\documentclass{interact}
\usepackage{siunitx}
\usepackage{tikz} 
\usepackage{graphicx}
\usepackage[natbibapa]{apacite}
\usepackage{setspace} 
\usepackage{xcolor}
\usepackage{amsmath} 
\usepackage{hyperref}
\usetikzlibrary{positioning}
\newcommand{\alphahat}{\hat{\alpha}}
\newcommand{\betahat}{\hat{\beta}}
\newcommand{\bAlpha}{\boldsymbol{A}}
\newcommand{\balpha}{\boldsymbol{\alpha}}
\newcommand{\bbeta}{\boldsymbol{\beta}}
\newcommand{\boldm}{\boldsymbol{m}}
\newcommand{\bM}{\boldsymbol{M}}
\newcommand{\hatM}{\widehat{\boldsymbol{M}}}

\newcommand{\bc}{\boldsymbol{c}}
\newcommand{\bC}{\boldsymbol{C}}
\newcommand{\dm}{\mathrm{d{\bf m}}}
\newcommand{\dt}{\mathrm{dt}}
\newcommand{\tmin}{t_\mathrm{min}}
\newcommand{\tmax}{t_\mathrm{max}}
\newcommand{\defint}{\int_{\tmin}^{\tmax}}
\newcommand{\TE}{\mathrm{TE}}
\newcommand{\NDE}{\mathrm{NDE}}
\newcommand{\NIE}{\mathrm{NIE}}
\newcommand{\DE}{\mathrm{DE}}
\newcommand{\IE}{\mathrm{IE}}
\newcommand{\xast}{ {x^\ast}} 
\newcommand{\changed}{\color{red}}

\begin{document}

\title{A causal approach to functional mediation analysis with application to a smoking cessation intervention}

 \author{
\centering\name{Donna~L.~Coffman\textsuperscript{$^\ast$}\thanks{CONTACT D.~L. Coffman. Email: dcoffman@mailbox.sc.edu}~PhD}
 \affil{Department of Epidemiology and Biostatistics, Temple University, Philadelphia, PA, 19122, USA}
 \centering\name{John~J.~Dziak~PhD}
 \affil{Edna Bennett Pierce Prevention Research Center, The Pennsylvania State University, University Park, PA, 16802, USA}
 \centering\name{Kaylee~Litson~PhD}
\affil{Instructional Technology \& Learning Sciences Department, Utah State University, Logan, UT, USA}
\centering\name{Yajnaseni~Chakraborti~MS}
\affil{Department of Epidemiology and Biostatistics, Temple University, Philadelphia, PA, 19122, USA}
\centering\name{Megan~E.~Piper~PhD}
\affil{Center for Tobacco Research Intervention, University of Wisconsin, Madison, WI, USA}
\centering\name{Runze~Li~PhD} 
\affil{Department of Statistics, The Pennsylvania State University, University Park, PA, 16802, USA}
}  
\maketitle
This article has been accepted for publication in Multivariate Behavioral Research, published by Taylor \& Francis. Dr. Coffman is now at the University of South Carolina.

\clearpage

\begin{abstract}
The increase in the use of mobile and wearable devices now allows dense assessment of mediating processes over time. For example, a pharmacological intervention may have an effect on smoking cessation via reductions in momentary withdrawal symptoms. We define and identify the causal direct and indirect effects in terms of potential outcomes on the mean difference and odds ratio scales, and present a method for estimating and testing the indirect effect of a randomized treatment on a distal binary variable as mediated by the nonparametric trajectory of an intensively measured longitudinal variable (e.g., from ecological momentary assessment). Coverage of a bootstrap test for the indirect effect is demonstrated via simulation. An empirical example is presented based on estimating later smoking abstinence from patterns of craving during smoking cessation treatment. We provide an R package, \verb|funmediation|, available on CRAN at 
\url{https://cran.r-project.org/web/packages/funmediation/index.html}, to conveniently apply this technique. We conclude by discussing possible extensions to multiple mediators and directions for future research.  
\end{abstract}

\vspace{40pt}


\vspace{50pt}
 
\clearpage

\section{Introduction}

Increasingly, researchers collect intensive longitudinal data on variables which are hypothesized to mediate the effect of an intervention on an end-of-study outcome of interest. For example, in a smoking cessation intervention study, individuals may be randomly assigned to a smoking cessation treatment such as the nicotine patch or varenicline. The first several weeks post-quit day are critical to maintaining smoking abstinence. Withdrawal symptoms (e.g., cravings, negative affect) may fluctuate even within a single day during these early weeks, and these fluctuations may have an effect on maintaining smoking abstinence. An effective smoking cessation treatment may have an effect on smoking abstinence that is due in part to changes in these mediating variables (i.e., due to a reduction in withdrawal symptoms).

Intensive longitudinal data can be collected using ecological momentary assessment \citep[EMA; see][]{Shiffman2008}, in which individuals are contacted multiple times per day for several weeks, and asked to report their cravings, affect, and other variables, including use behavior. The data need not be collected using EMA prompts; for example, it may be collected automatically with a mobile or wearable device. EMA data are also collected in other substance use cessation studies, and in weight loss and physical activity studies. In these studies, individuals may be assigned to a behavioral intervention, mediators are measured intensively over time, and the ultimate outcome of interest is whether the behavior change in question has successfully been maintained. The goal of this manuscript is to present and estimate a mediation model to evaluate the process by which a randomized intervention affects an end-of-study outcome via an intensively measured longitudinal process.

Over the last several decades, mediation analysis has been extended from single time point measures of a treatment, $X$, mediator, $M$, and outcome, $Y$, to multiple mediators, and to treatment, mediators, and outcomes that are measured longitudinally \citep{MacKinnon2008}. More recently, the assumptions needed to define and identify causal effects in the context of mediation have received considerable attention. As mediation is inherently about causal pathways, in which a treatment affects change in a mediator, which in turn affects change in the outcome, the potential outcomes framework has been employed to define and identify causal mediation effects \citep{VanderWeele2015}. 

Only in the last decade has mediation with an intensively measured treatment, mediator, or outcome within the potential outcomes framework been introduced. Specifically, \citet{Lindquist2012} proposed an approach for a mediator measured intensively by functional magnetic resonance imaging (fMRI).
Other advancements have included time-varying treatment and time-varying mediators \citep{VanderWeele2017,Park,Huang2017}, time-varying mediators and outcomes \citep{ZengFanLi2021,Loh2020,Huang2017},  time-varying treatment and mediator with a survival outcome \citep{Vansteelandt2019,Zheng}, and an exogenous (e.g., environmental) time-varying treatment with time-varying mediator and outcome \citep{Bind}. Additionally, \citet{Qian2019} study the mediation of a short-term effect of a time-varying treatment \citep[see also][]{KuanLiu2020}. However, here we focus on a single-time distal effect of a single-time treatment, as mediated by a process occurring between those times.

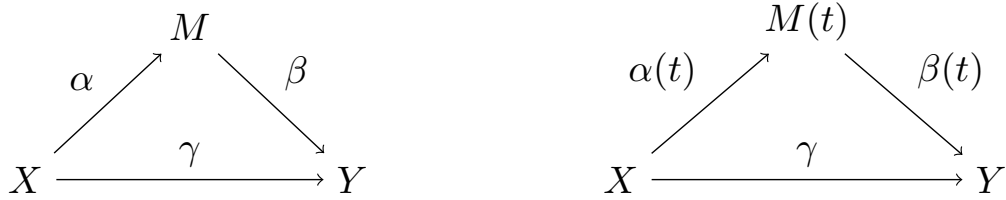
\begin{figure}[t] 
\centering 
\resizebox{\textwidth}{!}{
\begin{tabular}{ccc} 
\begin{tikzpicture}
\node (X) {$X$};
\node (M) [above right=of X] {$M$};
\node (Y) [below right=of M] {$Y$};
\path[->] (X) edge[auto] node{$\alpha$} (M);
\path[->] (M) edge[auto] node{$\beta$} (Y);
\path[->] (X) edge[auto] node{$\gamma$} (Y);
\end{tikzpicture}  &
\hspace{.5in} &
\begin{tikzpicture}
\node (X) {$X$};
\node (M) [above right=of X] {$M(t)$};
\node (Y) [below right=of M] {$Y$};
\path[->] (X) edge[auto] node{$\alpha(t)$} (M);
\path[->] (M) edge[auto] node{$\beta(t)$} (Y);
\path[->] (X) edge[auto] node{$\gamma$} (Y);
\end{tikzpicture} 
\end{tabular}}
\caption{ \large Simple (left) versus functional (right) mediation with single-time treatment and outcome.}
\label{fig:CrossSectionalVsFunctional}
\end{figure}

Specifically, we consider a time-invariant treatment, intensively measured mediator, and end-of-study outcome (see Figure \ref{fig:CrossSectionalVsFunctional}). This structure most closely resembles that of \citet{Lindquist2012}, but he considered only a continuous normally distributed outcome. We propose an extension to a binary outcome, such as smoking abstinence at follow-up in our motivating study. A binary outcome in a causal mediation analysis can be somewhat more complicated, even for a mediator measured at only a single time point \citep{RijnhartTwisk2017, RijnhartTwisk2019, RijnhartValente2021}. Causal effects for numerical outcomes are generally stated as differences in expected values. For binary outcomes, the difference in expected values is interpreted as a difference in probabilities (i.e., risk difference). However, if a risk ratio, odds ratio, or other quantity is desired, then causal mediation effects are defined differently \citep{VanderWeeleVansteelandt2010}.

Although \citet{VanderWeele2017} consider a single time point measurement of a binary outcome and a time-varying mediator, their approach only partially applies in our context because we, like \citet{Lindquist2012} and \citet{ZengFanLi2021}, view the mediation process as functional. One approach to fitting a mediation model in a setting with intensive longitudinal measurements is to replace a longitudinal process, $M(t)$ and/or $Y(t)$, with summary statistics such as means, sums, or estimated linear slopes.  This parametric summary approach could be made  sophisticated by treating  intercepts and slopes as subject-level latent random variables  \citep{Magidson2015,DeStavola2021,Sullivan2021}. However, as \citet{Lindquist2012} pointed out, it is also possible to take a less parametric approach, avoiding imposing linear shapes on trajectories, using ideas from functional regression \citep{RamsaySilverman2005}. Although \citet{ZengFanLi2021} propose functional mediation for an intensively measured mediator and intensively measured outcome, they do not propose a method for evaluating an intensively measured mediator with an end of study outcome. The functional mediation model proposed in this paper uses an outcome measured at a single point in time and thus, some of the assumptions made by \citet{ZengFanLi2021} are not necessary in the present context. Nevertheless, \citet{Lindquist2012}, \citet{VanderWeele2017}, and \citet{ZengFanLi2021} provide foundational ideas to the approach proposed here.



This article is organized as follows. In Section \ref{Sec:Example}, we present our motivating example. In Section \ref{Sec:Causal}, we introduce the notation, definitions, and identification assumptions needed in order for the effects to be interpreted as causal. In Section \ref{Sec:Est}, we present estimation of the indirect effect in detail, which we have implemented in an R package, \texttt{funmediation}. In Section \ref{Sec:Sim}, we present a simulation study to assess the coverage and power in data simulated to resemble to our real-data example. In Section \ref{Sec:Analysis}, we analyze the real data using the proposed model. 
Finally, in Section \ref{Sec:Discussion}, we discuss limitations, future directions, and conclusions.  In Appendix 1, we review the proof by \citet{Lindquist2012} for the identity link function and extend it to a log link function modeling a probability ratio. The more commonly used logistic link can be approximated by a log link in the case of rare outcomes \citep[see][]{Feingold2019,VanderWeeleVansteelandt2010,RijnhartValente2021}. 

\section{Motivating Example} \label{Sec:Example}

Our motivating example comes from the Wisconsin Smokers' Health Study 2 \citep[WSHS2; ][ClinicalTrials.gov identifier: NCT01553084]{Baker2016}. The WSHS2 is a comparative efficacy trial directly comparing the two most effective smoking cessation therapies (varenicline and combination nicotine replacement therapy [cNRT]) with one another and with an active comparator treatment (nicotine patch) \citep[see][]{Baker2016,LeFoll}.

Smokers ($n = 1086$; 52\% women, 67\% white, mean age = 48, mean number of cigarettes smoked/day = 17), recruited from Madison and Milwaukee, were randomly assigned to one of the three pharmacotherapies. The outcome considered here is carbon-monoxide (CO)-confirmed self-reported 7-day point-prevalence abstinence at 4 weeks post-target quit day (TQD). Participants completed one morning EMA prompt, one random prompt during the day, and one evening prompt every day for the first 2 weeks post-TQD and then every other day for the remaining 2 weeks of the EMA period. Thus, there are up to 66 EMA measurements that are not equally spaced due to fluctuations in the timing of the prompts and to sporadic missingness. Our proposed approach allows for these irregularly spaced assessments. The goal of our empirical analysis is to examine whether the intervention of interest (varenicline) has an effect on smoking cessation (7-day point-prevalence abstinence at 4 weeks post-TQD) that is mediated by a reduction in craving, compared to any nicotine replacement therapy. Craving is an important predictor of relapse in many substance use disorders \citep{Vafaie,LeFoll}, and it was assessed at each EMA prompt during the 4 week post-TQD period. It is treated as a time-varying mediator in Section \ref{Sec:Analysis}.

\section{Proposed Approach} \label{Sec:Causal}

\subsection{Notation}

Consider a sample of $N$ individuals, in which each individual $i, i=1, \ldots, N$ is randomly assigned to one of two treatments, $X_i=0$, the control or comparison group, or $X_i=1$, the active treatment group. Note that this notation is easily extended to more than two treatment conditions. Each individual $i$ is assessed at $T_i$ measurement occasions, which may vary across individuals, between times $\tmin$ and $\tmax$. At each measurement occasion, $t_{ij}, j = 1, \ldots, T_i$, such that $\tmin \le t_1 \le \ldots \le t_{T_i}$, the mediator, $M_{i}(t_{ij})$ is assessed. The end-of-study outcome, denoted $Y_i$, is measured at a distal time point, which is after $\tmax$ and common to all individuals. Let $\bM_i = \{ \bM_i(t): t \in [\tmin,\tmax]  \} $ denote the complete mediator process for individual $i$ from beginning to end of the study interval, including times at which it is not  observed.  
In the context of the potential outcomes framework, we additionally use parentheses to denote the counterfactual outcomes under each treatment condition. Specifically, let $\bM_i(x)$ be the potential value that the complete mediator process would take for individual $i$, if that individual, perhaps counterfactually, received treatment level $x$. Let $Y_i(x, \bM_i(\xast))$ be the potential outcome if individual $i$, perhaps counterfactually, is given treatment $X=x$ and mediator process $\bM_i(\xast)$.

We make the stable unit treatment value assumption \citep[SUTVA;][]{Rubin1980} of non-interference between individuals. That is, an individual's potential values on the mediator process are not dependent on any other individual's treatment assignment and an individual's potential outcomes are not dependent on any other individual's treatment assignment or potential values of their mediator process. SUTVA also states that there are not multiple versions of the treatment \citep[see][]{HernanWhatIf}. In addition, we make a consistency assumption, in which the counterfactual outcomes $Y_i(x)$ and $\bM_i(x)$ are equal to the observed outcomes, $Y_i$ and $\bM_i$, respectively, when individual $i$ is assigned to treatment level $x$, and the composition assumption, i.e., $Y_i(x) = Y_i(x, \bM_i(x))$ \citep{VanderWeeleVansteelandt2009,VanderWeeleVansteelandt2010,Lindquist2012}. Finally, let $Y_i(x,\boldm)$ denote the potential outcome for individual $i$ given treatment level $x$ and given mediator process, $\boldm$. That is, $\bM_i(x) = \boldm$, a specific mediator process. Where possible, to avoid confusion, we drop the $i$ subscript for simplicity.

\subsection{Definitions}

In this paper we focus on the natural direct and indirect effects, similar to \citet{VanderWeeleVansteelandt2010} and \citet{RijnhartValente2021}. In contrast, \citet{VanderWeele2017} and \citet{PintoPereira2020}, focus on the interventional direct and indirect effects; for an explication of differences between these definitions, see \cite{Moreno2018} and \cite{Nguyen2021clarifying}.

If an identity link is used, the causal effect is most readily expressed in terms of a mean difference. In this case, the average causal total effect of the treatment on the distal end-of-study outcome is defined as:
\begin{equation*}
    \TE \equiv E(Y(1,\bM(1)))- E(Y(0,\bM(0))) =
    E(Y(1))- E(Y(0))
\end{equation*}
which can be decomposed into the natural direct effect:
\begin{equation*}
\NDE(0) = E(Y(1,\bM(0)))-E(Y(0,\bM(0))) 
\end{equation*}
and the natural indirect effect:
\begin{equation*}
\NIE(1) = E(Y(1,\bM(1)))-E(Y(1,\bM(0)))
\end{equation*}
obtained by adding and subtracting $E(Y(1,\bM(0)))$. The natural direct effect, $\NDE(0)$, defines the effect of the treatment on the outcome holding the mediator process to the potential value obtained under the control condition (i.e., $x=0$). The natural indirect effect, $\NIE(1)$, defines the effect on the outcome of the potential mediator process under treatment vs. the potential mediator process under control while holding treatment level fixed to treated (i.e., $x=1$) \citep[see, e.g.][]{VanderWeele2015}. Alternatively, one could add and subtract $E(Y(0,\bM(1)))$ to obtain an equivalent decomposition: $\NDE(1) = E(Y(1,\bM(1)))-E(Y(0,\bM(1)))$ and $\NIE(0) = E(Y(0,\bM(1)))-E(Y(0,\bM(0)))$. We use the former decomposition, $\TE = \NDE(0) + \NIE(1)$, although the procedures are applicable to the latter as well. 

For modeling binary or count data, a log or logistic link may be used instead of an identity link function. That is,
\begin{align*}  
g(\TE) =&  g\big( E(Y(1,\bM(1)))\big)- g\big(E(Y(0,\bM(0)))\big)\\
g(\NDE(0)) =&  g\big(E(Y(1,\bM(0)))\big)-g\big(E(Y(0,\bM(0)))\big) \\
g(\NIE(1))  =&  g\big(E(Y(1,\bM(1)))\big)-g\big(E(Y(1,\bM(0)))\big)   
\end{align*}
where $g$ represents the log or logit link. In this case, the causal effect definitions are on the ratio scale (either rate ratio for the log link or odds ratio for the logit link). For the log link this would be:
\begin{align}
\label{eqnLogLinearEffects}
    \TE =&  E\big(Y(1,\bM(1))\big) / E\big(Y(0,\bM(0))\big) \\
    \NDE(0) =&  E\big(Y(1,\bM(0))\big) / E\big(Y(0,\bM(0))\big) \nonumber \\ 
    \NIE(1) =&  E\big(Y(1,\bM(1))\big) / E\big(Y(1,\bM(0))\big). \nonumber 
\end{align}
Equivalently, the log of each effect is a difference of log probabilities or log odds, respectively.

\subsection{Identification Assumptions}\label{Sec:Assumptions}

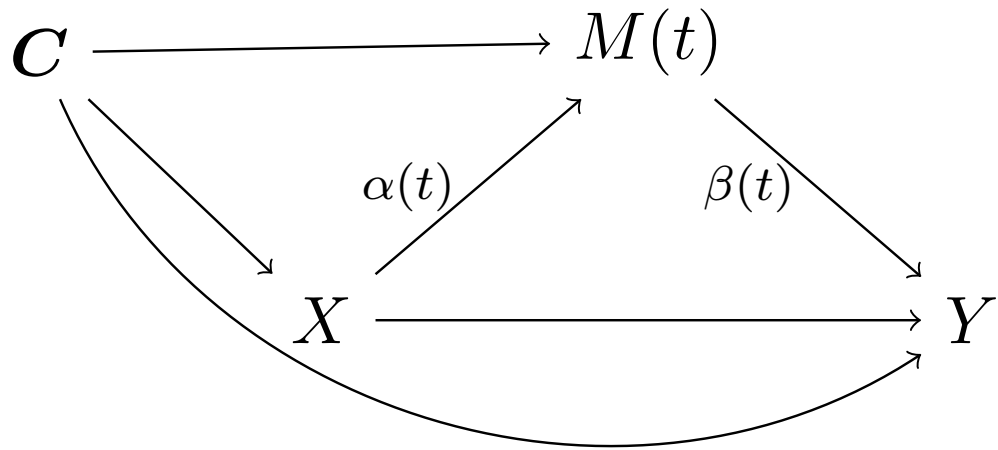
\begin{figure}[t] 
\resizebox{\textwidth}{!}{
\begin{tikzpicture}
\node (X) {$X$};
\node (C) [above left =of X] {$\boldsymbol{C}$};
\node (M) [above right=of X] {$M(t)$}; 
\node (Y) [below right=of M] {$Y$};
\node(secret) [above=of M] {};
\path[->] (C) edge[auto] (X);
\path[->] (C) edge[auto] (M); 
\path[->] (C) edge[bend right=50] (Y);

\path[->] (X) edge[auto] node[left]{\scriptsize{$\alpha(t)$}} (M);
\path[->] (M) edge[auto] node[left]{\scriptsize{$\beta(t)$}} (Y);
\path[->] (X) edge[auto] (Y);
\end{tikzpicture} 
}
\caption{ \large Mediation model with functional $M$, including baseline covariates $\boldsymbol{C}$.}
\label{fig:FunctionalWithCovariates}
\end{figure}

In addition to the aforementioned SUTVA, consistency, and composition assumptions, identifying the potential outcomes from the observed data requires the following ignorability assumptions given baseline covariates, $\bC$, as shown in Figure \ref{fig:FunctionalWithCovariates}. In addition, we require the positivity assumption which states that each individual has a positive probability of being assigned to either treatment and to a given mediator process given their covariate values, $P(X=x|\boldsymbol{C}=\boldsymbol{c}) > 0$ and $P(\bM_i(x)=\boldsymbol{m} |X = x, \boldsymbol{C}=\boldsymbol{c}) > 0$. The assumptions allow the identification of the potential outcomes from the observed data. The assumptions are adapted from  \citet{VanderWeeleVansteelandt2010} \citep[see also][]{Imai2010a}, but the first three are analogous to Assumption 1 of \citet{Lindquist2012}, and the last is analogous to Assumption 2 of \citet{Lindquist2012}. 

\begin{enumerate}
    \item Conditional on $\bC$, the effect of the treatment on the outcome is unconfounded (i.e., there are no unmeasured confounders of the treatment and outcome): $Y_i(x, \boldm) \perp X_i | \boldsymbol{C}_i$.
    \item Conditional on $X$ and $\bC$, the effect of the mediator on the outcome is unconfounded (i.e., there are no unmeasured confounders of the mediator and outcome): $Y_i(x, \boldm) \perp \bM_i | X_i, \boldsymbol{C}_i$.
    \item Conditional on $\bC$, the effect of the treatment on the mediator is unconfounded (i.e., there are no unmeasured confounders of the treatment and mediator): $\bM_i(x) \perp X_i | \boldsymbol{C}_i$.
    \item  There are no measured or unmeasured confounders of the mediator and outcome that have been influenced by the treatment: $Y_i(x, \boldm) \perp \bM_i(\xast)| \bC_i$ for all $x$ and $\xast$.
\end{enumerate}

Assumptions 1-3 are standard sequential ignorability assumptions \citep{Imai2010a}. Given randomization to the treatment levels, the first and third assumptions are reasonable. In contrast, Assumption 2 may not hold even when there is randomization to the treatment levels, because there is not randomization to the mediating process; however, if a rich set of potential confounders have been measured, then this assumption may still be reasonable conditional on the measured confounders. 

Assumption 4 (often referred to as cross-world independence) may be problematic. Although the outcome occurs at a distal time point in our context, and therefore is not itself a time-varying confounder \citep[e.g., as in][]{ZengFanLi2021}, there may be other potential mediating processes which may violate this assumption. For the present motivating example, other variables were measured during the EMA period, such as smoking cessation self-efficacy, self-reported daily smoking, and smoking cessation fatigue. It may be expected that the treatment may influence these variables, which in turn may affect both the mediating process of interest (i.e., craving) and the outcome, (i.e., abstinence at one month post-TQD). We further discuss possibilities for allowing these additional mediators in Appendix 2.

There are possible alternative assumptions to Assumption 4 when it may be violated \citep[see, e.g.][]{VanderWeele2015, Robins2003}. Rather than assuming that there are no mediator-outcome confounders affected by treatment (i.e., Assumption 4), an alternative identifying assumption is to assume that there are no interactions between treatment assignment and the mediating process. However, this assumption is not as innocuous as it may seem \citep{Robins2003}. It means that the natural direct and indirect effects are the same in both the treatment and control groups (i.e., $\NDE(0) = \NDE(1)$ and $\NIE(0) = \NIE(1)$). In this case the controlled direct effect, defined as $Y(1, \boldm)-Y(0, \boldm)$ for some specified fixed trajectory $\boldm$ for the mediator process, (i.e., $\bM (x) = \boldm$), is equal for any level of $\boldm$ and is also equal to the $\NDE$. Then the $\NIE$ can be parametrically identified as $\TE-\NDE$, as in the traditional linear structural equation modeling approach to mediation. However, this lack of interaction may also be difficult to justify. Note that \cite{ZengFanLi2021}, \citet{Lindquist2012}, and \citet{zhao2018functional} all impose Assumption 4 or the no-interaction assumption. 

\citet{Lindquist2012} discusses an alternative approach to estimating the causal effect of the mediator on the outcome, without making Assumption 4. He proposes the intuition of using the ratio of the total effect of $X$ on $Y$ to the effect of $X$ on $M$ as a measure of the strength of mediation through $M$. This could be described as using $X$ as an instrumental variable for the purposes of the causal effect of $M$ on $Y$. However, this alternative approach requires an exclusion restriction assumption (no direct effect of $X$ on $Y$, i.e., complete mediation) which will often be unrealistic. In the current example, it does not seem plausible to assume that reduced craving is the only mechanism by which treatment affects relapse probability.

In any case, identifying the natural direct and indirect effects from the observed data requires strong assumptions. If Assumption 4, the no-interaction assumption, or the exclusion restriction are not realistic, then it may be possible to define interventional direct and indirect effects \citep[see][]{VanderWeele2017}, rather than the natural direct and indirect effects considered here. This approach is rather new, and applying it to functional regression requires further methodological research, but it is a promising future direction.

Given Assumptions 1 - 3 and the no-interaction assumption (or Assumption 4), the natural direct and indirect effects can be identified from the observed data. The causal effects can be estimated as functions of the expected values (e.g., $E[Y|X, \boldsymbol{C}, \bM)$) of the observed data. Next, we describe the models we use to estimate these quantities.



\section{Estimation} \label{Sec:Est}

In general, the estimation strategy is a two-step process in which first a model is fit to predict the mediator from treatment and baseline covariates, and then another model is fit to predict the outcome from the mediator, treatment, and baseline covariates. 

\subsection{Model for Mediator}

For the identity link, the model for the potential values of the mediator at time $t$ for individual $i$ if given treatment $x$ is:
\begin{equation}
\label{eqn:M_subject}
  M_i(t; x) = \alpha_0(t) +  \bC_i \balpha_C(t)  +  \alpha_X(t)x + \epsilon_{i}(t)
\end{equation} 
where $\epsilon_i$ is a random process representing  measurement-level errors for the mediator, which are assumed to have $E(\epsilon_{it})=0$ for all $t$ and may be correlated within person, and where $\alpha_0(t)$,  $\balpha_C(t)$, and $\alpha_X(t)$ are smooth functions of time. By using the consistency assumption, we can simply write the potential value, $M_i(t;x)$ which is a function of both $t$ and treatment level $x$, as $M(t)$ when expressing an expected value conditional on $x$.

More generally, we can model $M(t)$ as 
\begin{equation}
\label{eqn:M}
    g(E(M(t) | X=x, \bC)) = \alpha_0(t) +
   \bC  \balpha_C(t) +     \alpha_X(t) X
\end{equation}
where $g$ is a link function such as log, logit, or identity. 

Model (\ref{eqn:M_subject}) or (\ref{eqn:M}) is an application of concurrent function-on-scalar regression \citep{RamsaySilverman2005} or equivalently the time-varying effect model \citep[TVEM;][]{Hastie1993,Tan2012}. The link function $g$ is treated as given and known, but the $\alpha$ functions are not assumed to have any simple parametric form known to the data analyst.

\subsection{Model for Outcome}

The model for the potential outcome using an identity link is given as: 
\begin{equation}
\label{eqn:Y_subject}
  Y_i(x_i, \bM_i(x_i)) = \beta_0 + 
 \bC_i \bbeta_C  +  
  \beta_X x_i + \defint \beta_M(t) \bM_i (t;x) \dt  + \eta_i,
\end{equation} 
where the $\eta$ are subject-level errors for the response assumed to have  $E(\eta|X=x, \bM(X)=\boldm)=0$. The subject-level errors $\eta_{i}$ for the outcome in (\ref{eqn:Y_subject}) are treated as independent of the measurement-level errors $\epsilon_i(t)$ for the mediator in (\ref{eqn:M_subject}).

In general, we can model the expected value of the observed outcome, $Y$, given treatment $x$, covariate vector $\bc$, and mediator trajectory $\boldm$, as
\begin{equation}
\label{eqn:Y}
g(E(Y|X=x, \bC=\bc, \bM=\boldm)) = \beta_0 + \beta_X x +
 \bc \bbeta_C  +  \defint \beta_M(t) \boldm(t) \dt
\end{equation}
where $g$ is a link function such as an identity, log, or logit. A model such as (\ref{eqn:Y_subject}) and (\ref{eqn:Y})  represents scalar-on-function regression \citep{RamsaySilverman2005, GoldsmithBobb2011, Lindquist2009}.

\subsection{Expression for the Natural Indirect Effect}

Using the identity link, and under Assumptions 1-3 and the no treatment-mediator interaction assumption, the $\NIE(0)$ and the $\NIE(1)$ are equal. Thus, the natural indirect effect is:
\begin{align*}
\NIE &= E(Y(0,\bM(1))-Y(0,\bM(0))) \\ 
&= E(Y(1,\bM(1))-Y(1,\bM(0)))\\
&= \defint \alpha_X(t) \beta_M(t)\dt
\end{align*}
\citet{Lindquist2012} proved this for the identity link.

For a binary outcome using a log link,
\begin{align*}
    \NIE &= E(Y(0,\bM(1)))/E(Y(0,\bM(0))) \\
&=E(Y(1,\bM(1)))/E(Y(1,\bM(0)))  \\
&=\exp \Big( \log \big(E(Y(0,\bM(1)))\big) - \log \big(E(Y(0,\bM(0)))\Big) \\
&= \exp  \Big( \defint \alpha_X(t) \beta_M(t) \dt \Big), 
\end{align*}
and the proof is given in Appendix 1. For binary data using a logistic link, it is more difficult to express the decomposition of the total effect into direct and indirect effects in a straightforward way, but the logit link function approximates a log link function when outcomes are rare \citep[see][]{VanderWeeleVansteelandt2010,RijnhartValente2021}. In any case, it remains reasonable to use a significance test for $\int \alpha_X(t) \beta_M(t) \dt$ as a test for significant mediation, by analogy to the product of coefficients method in traditional mediation analysis. Another option, not pursued here, would be to modify the definition of direct and indirect effects in order to make the logistic link easier to handle theoretically, following either the ideas in \citet{Geldhof2017} or \citet{Doretti2021}.

The controlled direct effect of $X$ on $Y$ given a fixed mediator trajectory  $\boldm$ is  $E(Y(1,\boldm))-E(Y(0,\boldm)) = \beta_X$ for the identity link or $E(Y(1,\boldm))/E(Y(0,\boldm)) = \exp(\beta_X)$ for the log link. Because of the assumption of no treatment-mediator interaction, the controlled direct effect of $X$ given $\bM(x)=\boldm$ equals both the $\NDE(0)$ and $\NDE(1)$. 

\subsection{Model for Total Effect}
 
Finally, the total effect using an identity link is given as follows:  
\begin{align*}
    \tau &\equiv E(Y(1,\bM(1))) - 
              E(Y(0,\bM(0))) \\
              &= \beta_X + \defint \alpha_X(t) \beta_M(t)\dt
\end{align*}
For the log link, the total effect is additive on a log scale, 
$$\exp (\tau) = 
\exp (\beta_X) + \exp \Big( \defint \alpha_X(t) \beta_M(t)\dt \Big),$$
but is multiplicative on the ratio scale. Again, a logit link is also possible, although it can make the relationship between direct, indirect, and total effects difficult to express unless the outcome is rare.

\subsection{Test for Indirect Effect}

To test for an indirect effect, one can test whether a nonparametric bootstrapped confidence interval for $\int \alpha_X(t) \beta_M(t) \dt $ includes zero \citep{Lindquist2012}. In this context, the indirect effect is a scalar, even though its individual effects, $\alpha_X(t)$ and $\beta_M(t)$, are functions of $t$. 

\subsection{Time-varying effect models}

Models (\ref{eqn:M_subject}) and (\ref{eqn:M}) can be fit as longitudinal varying-coefficients models, i.e., time-varying effect models \citep[TVEM; see][]{Tan2012}. In general, a TVEM model with a time-varying response variable, $M(t)$, is specified as 
$$g(E(M(t)))= \alpha_0(t) +\sum_j \alpha_j (t) X_j(t)$$
for a set of longitudinal predictors $X_j(t)$ observed on a sparse and possibly irregular grid. In this general version, both the covariates and their ``effects'' (time-specific regression coefficients) are time-varying. However, any of the $\alpha_j$ and/or $X_j$ can be optionally constrained to be constant over time.  

Notice that for parsimony and interpretability, and to avoid the curse of dimensionality, TVEM assumes a linear or generalized linear structure at any given time, and that the estimates of $\alpha_0$ and the $\alpha_j$'s are smooth but may potentially be nonlinear functions of time. Thus, Models (\ref{eqn:M_subject}) and (\ref{eqn:M}) are included as a special case. The TVEM model is straightforward to fit as a marginal model using working independence and robust standard errors, and a spline approximation to the functional coefficients. Thus, a TVEM marginal model is essentially a generalized estimating equations \citep{LiangZeger1986} model with some of the parametric terms replaced by splines. 

We fit the model by representing the functional coefficients as penalized B-splines \citep{EilersMarx1996}, as implemented in the \verb|tvem| function of the \verb|tvem| R package \citep{package_tvem}. The model is fitted approximately by first constructing a spline basis for the time interval of interest, and then performing a large parametric regression including each function of time in the basis, as well as its product with each coefficient. We use a penalty function on the nonlinear functions of time in order  to avoid overfitting and control  the level of smoothness versus flexibility of the fitted function.  By default, \verb|tvem| uses cubic B-splines with a maximum of 20 interior knots and a first-order difference penalty on the knot coefficients \citep[see][]{EilersMarx1996}.  Penalized truncated power splines \citep[see, e.g.,][]{Tan2012,Huang2017} are an alternative option. 

TVEMs can be viewed as a special case of the varying-coefficient model \citep{Hastie1993} in which the coefficients are indexed by time. The varying coefficients allow the effect of treatment on the mediator to change as a nonparametric function of time. Additional time-varying and time-invariant predictors can also be added to the model. Predictors can be specified to have either time-varying or time-invariant effects. For example, demographic variables or treatment group membership may have time-varying relationships with the outcome variable even if they are themselves not time-varying. In the \texttt{tvem} package, within-subject correlation is accounted for indirectly by using sandwich standard errors, similar to working-independence GEE \citep{LiangZeger1986}. This can be seen as an application of penalized pseudo-likelihood, incorporating the penalty required to regularize the spline functions. The penalty function is tuned by a restricted maximum likelihood (REML) approach (Wood, 2001). This approach shrinks the changes at each knot, acting analogously to an empirical Bayes prior. The prior distribution used in the \texttt{tvem} package is a first- or second-order difference penalty as described in Eilers \& Marx (1996), and it acts to shrink the fitted curve towards a straight line (possibly even a flat line with zero slope) unless data suggests otherwise.

\subsection{Scalar-on-function regression}

Although the underlying process, $\bM_i$, is assumed to be a continuous function, practically it is observed at only a finite number of points and potentially has measurement error. For computational purposes, it would be very helpful to have values for $M_i(t)$ on a dense, regular grid without missing values. However, this is not likely to be the case in practice. One option for working around this difficulty, if the amount of missing data is not too great, is by linear interpolation between nonmissing points. Another option is to use spline regression to fit a smoothed curve $\hatM_i$ to the observed data for each participant $i$, and then using these smoothed values to estimate the $\beta_M(t)$ function. This smoothing could be heuristically justified by supposing that the mediator is measured with error, and that the true trajectory of the mediator was smoother than the trajectory which would have been recorded if all possible measurement occasions on the grid had been observed for all participants. That is, we let $\widehat{M}_i(t)$ at time $t_{ij}$ be a possibly smoothed prediction of the true latent curve underlying $M_i(t)$, rather than the observed (and presumably error-contaminated) $M_i(t)$ itself \citep{GoldsmithBobb2011} by incorporating a presmoothing step in the model (see Appendix 1). 

After this presmoothing, the fitted model is 
\begin{equation}
    \label{eqn3}
g(E(Y_{i}|\bM_{i}(t))\approx\beta_{0}+
\bbeta_C \bC_i + \beta_X X_i +
\intop_{\tmin}^{\tmax}\beta_{M}(t)
\widehat{M}_{i}(t) dt.    
\end{equation}

Additional scalar (i.e., time-invariant) predictor variables with scalar coefficients may be added to Model (\ref{eqn3}). The coefficients, both scalar and functional, are estimated following the general method of \citet{GoldsmithBobb2011} using REML. The coefficient functions are represented via the penalized thin plate regression splines approach described by \citet{Wood2003} and \citet{mgcvBook}. Notice that the presence of within-subject variability is required, because if $\bM_i(t)$ were constant over $t$ for each $i$, then $\beta_M(t)$ would not be identified. Accordingly, with substantial and meaningful within-subject variability, the functional regression coefficient, $\beta_M(t)$, requires regularization using a penalty function and/or a low-dimensional basis restriction.

\subsection{Overall indirect effect computation}

Once $\alphahat_X(t)$ and $\betahat_M(t)$ have been estimated, the indirect effect is operationally defined as the integral $\defint\alphahat_X(t)\betahat_M(t)dt$. This is approximated by taking the mean of $\alphahat_X(t)\betahat_M(t)$ for an evenly spaced grid of points $t$ and then rescaling by multiplying it by the width of the time interval of interest.

Next, a nonparametric bootstrap significance test for an overall $p$-value for $\defint\alpha_X(t)\beta_M(t) \dt$ is obtained. A simple approximate confidence interval for this overall quantity using either a normal approximation or the percentile method can be obtained using the \texttt{funmediation} package.

\subsection{Difficulties with time-specific indirect effects}

We focus on an overall test of mediation, rather than time-specific mediation estimates, and do not attempt to interpret $\alpha_X(t)\beta_M(t)$ at a specific $t$. This is in contrast to \citet{Lindquist2012} who used a ``wild bootstrap'' to construct confidence intervals for the point-specific indirect effect $\alpha_X(t)\beta_M(t)$ at each of a grid of points $t$. However, the wild bootstrap is more difficult to apply to binary outcomes, since it involves permutation of residuals. Also, the practical meaning of the time-specific indirect effect $\alpha_X(t)\beta_M(t)$ with EMA data could be more difficult to interpret because of temporal dependence  \citep[see][]{Dziak2019,VanderWeele2017} and because time-specific snapshots may not adequately represent time-lagged effects \citep[see][]{Maxwell2011}; we address this further in the Discussion.


\section{Simulations} \label{Sec:Sim}

We performed a simulation study to determine whether the proposed method has adequate coverage and power under realistic scenarios. We simulated 5000 datasets for each of three sample size scenarios ($N=250$, $N=500$, and $N=1000$). In each dataset, the treatment $X$ is dichotomous and generated with equal probability of 0 (representing control group) and 1 (representing treated group). We imagine an interval from 0 to 1 representing rescaled time in a smoking cessation program, $X$ as a randomized intervention to reduce cravings, $M_i(t)$ as recent amount of cravings reported at time $t$, and $Y$ as relapse status by end of program, with $Y=1$ indicating relapse. These simulated variables map onto the applied example, but the specific values of the simulated variables represent a much wider array of possible variable combinations.

\subsection{Methods}

The mediator is generated on a grid of 100 time points on an interval from 0 to 1 (e.g., 0.00, 0.01, ..., 0.99). However, 60\% of those observations are randomly deleted, to create uneven person-specific measurement times. $\bM_i$ as a whole is assumed to follow a Gaussian process with mean function $E(M_i(t) | X_i)=\alpha_0(t) + \alpha_X(t) X_i$. We set the time-varying mean of the mediator for the control group to $\alpha_0(t) = \sqrt{t}$, and the time-varying effect of $X$ on the mediator to $\alpha_X(t)=-\sqrt{t/2}$. This means that both the control and treatment groups start on average with a mean near 0 on the mediator, but the control group increases nonlinearly over time to near $1$, while the treatment group increases nonlinearly to near $0.3$. The error process for the mediator is generated to have a standard deviation of $2$ and an AR(1) structure with correlation $0.8$ between adjacent measurements on the full 100-point grid.  

The outcome $Y_i$ was generated as binary, with a logistic link. Specifically, success probability was given by 
$$\mathrm{logit}(E(Y_i)) = \beta_0 + \beta_X X_i + \int_0^1 \beta_M (t) M_i(t) \dt$$ 
where $\beta_0 = 0$, $\beta_X = 0.2$, and $\beta_M = \frac{1}{2} (\exp(t)-1)$. Therefore, $X$ tends to reduce $M(t)$, and $M(t)$ is positively related to $Y$, such that the overall indirect effect is negative (a negative coefficient times a positive coefficient). Specifically, the overall true value of the indirect effect is taken to be $\int_0^1 \alpha_X (t) \beta_M (t) = -0.2113$. Because the treatment is simulated to be beneficial and the simulated outcome is imagined as deleterious, the negative effect is reasonable.

The model was fit to each dataset using the proposed combination of TVEM and scalar-on-function regression. We tested the null hypothesis that the indirect effect was zero using 199 bootstrap replications. In a real data analysis, 199 would be inadequate, and at least 999 would be recommended. However, we found that the larger number was computationally infeasible in a simulation setting, because 1000 simulated datasets per scenario with 999 bootstrap replications would mean having to fit and tune the model separately for each of approximately one million datasets per scenario.

The R software environment \citep{RManual} was used, with author-written functions available in the package \verb|funmediation| available on CRAN. This package employs the \verb|tvem| package \citep{package_tvem} and the \verb|pfr| function in the \verb|refund| package \citep{package_refund}. Each package uses generalized additive model functionality from the \verb|gam| or \verb|bam| functions in the \verb|mgcv| package \citep{mgcvBook,package_mgcv} for back-end calculations.  Presmoothing is  handled via interpolation using the \verb|presmooth| option in the \verb|lf| function in the \verb|refund| R package. Bootstrapping is performed using the \verb|boot| function in the \verb|boot| package \citep{boot1,boot2}.

\subsection{Results}

Performance measures for estimates of the relationship of $X$ to $M(t)$ are shown in Table \ref{Table:Results:XM}. Estimates for $\alpha_0(t)$ and $\alpha_X(t)$ had very little bias and had increasing precision as sample size increased.   

Performance measures for estimates of the relationship of $X$ and $M(t)$ to $Y$ are shown in Table \ref{Table:Results:MY}. Results indicate a very small bias of the estimated effects towards zero, which does not seem to vanish with increasing sample size. Nonetheless, random error decreases with increasing sample size, and pointwise coverage for the $95\%$ pointwise confidence intervals is essentially nominal for all sample sizes, so performance can generally be considered good. The small but non-vanishing bias present may be a side effect of the shrinkage penalty functions used to prevent overfitting of $\alpha_X$ and $\beta_M$.

The confidence intervals for $\beta_M(t)$ are constructed only for pointwise coverage (i.e., a 95\% long-term probability of containing the true value at any fixed time point), not nominal familywise coverage (i.e., a 95\% long-term probability of covering the true value for all the time points jointly), so they would not perform as well on familywise coverage.   \citet{Lindquist2012} recommended false discovery rate correction over values of $t$. However, in the case of observational studies with potentially diffuse and weak effects over time, this might lead to a very high Type II error rate. That is, researchers might misinterpret points which are indicated as significant at a given $\alpha$ level as being fundamentally different from time points indicated as nonsignificant, i.e., as times when the exposure or treatment are active and inactive, respectively, despite both being highly affected by sample size and sampling error. It is not yet known how best to handle this issue. Fortunately, it is not a problem for the overall indirect effect $\int_0^1 \alpha_X \beta_M \dt$, because it is a scalar and only tested once.

Performance measures for estimates of the indirect relationship of $X$ to $Y$ are shown in Table \ref{Table:Results:XY}. There was a small conservative bias towards zero, which became even smaller for larger sample sizes. Mean squared error also decreased for larger sample sizes. Power and coverage are shown for the normal-theory bootstrap, basic bootstrap, and percentile bootstrap. Of the three methods, the best coverage and power are obtained using the percentile bootstrap. Coverage is very good, and power increases with larger sample sizes. However, adequate power requires more than 500 participants.

\section{Empirical Analysis} \label{Sec:Analysis}

In this section, we present the results from analyzing the data described in Section \ref{Sec:Example}. The R code for implementing this analysis is available in the supplementary materials.

\subsection{Sample Characteristics and Missing Data}

Our analysis sample, taken from the 1086 participants in the WSHS2 \citep{Baker2016}, consisted of 1041 participants who were randomized, provided data to at least one EMA prompt post-TQD, and whose abstinence or non-abstinence status, $Y$, could be ascertained. There was very little missingness for $Y$, which was not subject to noticeable differential missingness across treatment group (11 missing $Y$ values from among those randomized to patch group, 13 from the varenicline group, 11 from the cNRT group). Thus, missingness is not a large concern for the outcome, $Y$.
 
However, participants varied widely in the proportion of response to the total 66 EMA prompts possible, so the mediator $M(t)$ at any given time $t$ was subject to noticeable missingness. This may be a concern for assessing the relationship between $X$ and $M(t)$ or between $M(t)$ and $Y$.

Among the 1086 participants, all of whom have $X$ (randomization status), 45 did not provide any EMA data, and they were excluded from the analysis. The 1041 participants who were included provided between 1 and 66 observations per participant (mean=41, median=45). There were three EMA prompts per day: morning, evening, and a random time in between. For participants who responded to most prompts, the middle, random, prompt was more likely to be missing than either the morning or evening prompts. The random prompt may have been missing because the participant merely happened to be temporarily unavailable to answer. This is likely to be less of a problem than long-term dropout.

Because the current analysis is intended mainly as an illustration of a new technique, we do not attempt to solve the possible problem of informative missingness here. However, the serious potential limitations which it poses, and possible future approaches to address them, are discussed in Section \ref{Sec:Discussion}. We conducted an additional analysis in which participants who responded to two or fewer EMA prompts post-TQD were excluded as a sensitivity analysis to partially assess the impact of missingness of functional data on estimation using this technique. The results are similar to those presented here and are included in Appendix 3.

\subsection{Total Effect}

For comparison, we fit a logistic regression model of the outcome, CO-confirmed 7-day point-prevalence abstinence (1=abstinent, 0=relapsed), at four-weeks post-TQD on treatment group (1 = varenicline, 0 = nicotine replacement therapy). The total effect estimate from this model was not statistically significant (estimate = 0.023, std. error = 0.135, $z = 0.167$, $p = 0.868$). Because this study does not have a no-treatment control group, and instead focuses on comparative efficacy between active treatments, this null finding may be due to the true difference between treatment efficacy being small (i.e., small effect size for comparing treatment levels). However, the example is still a useful illustration of the proposed procedures.

\subsection{Direct and Indirect Effects}

Next, we examine the direct and indirect effects of craving, defined as the desire to smoke in the last 15 minutes, throughout the first two weeks post-TQD using the proposed method. The results of these analyses are unique, providing insights that are not discernible from any other analytic approach, including other mediation approaches. The indirect effect estimate was statistically significant (estimate = 0.044, percentile method 95\% bootstrap confidence interval = (0.006, 0.111)). The estimate of the direct effect was not statistically significant (estimate = -0.067, std. error = 0.139). The indirect and direct effects do not add up exactly to the total; this is likely due to the effects of smoothing of the mediator and of the estimated effects (see Appendix 1). It is likely that oversmoothing or undersmoothing of the mediator trajectories may have an effect like that of measurement error, which is known to cause bias \citep[see][]{Fritz2016,DeStavola2021}. 

\subsection{Effect of Treatment on Mediator}

\begin{figure}[t]  
\includegraphics[width=5.6in,height=4in]{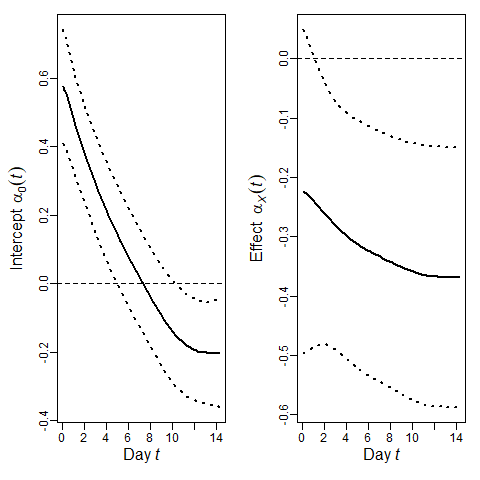} 
    \caption{ \large Time-varying effect of varenicline (vs. nicotine patch only) on the mediator, craving to smoke in the last 15 min. Left and right panes represent time-varying intercept function and time-varying effect of treatment, respectively. Upper and lower dotted lines represent pointwise approximate 95\% confidence intervals.}
    \label{fig:tvem} 
\end{figure}

Examination of the individual effects that comprise the indirect effect revealed a more nuanced interpretation. Figure \ref{fig:tvem} illustrates the time-varying effect of treatment on the mediator. The effect is such that varenicline reduces craving  and this effect becomes slightly stronger (i.e., more negative)  during the first week post-TQD before leveling off. Other than the first day post-TQD, the effect is statistically significant throughout the two-week post-TQD period, at least in the heuristic sense that the pointwise confidence intervals generally do not include zero.

\subsection{Effect of Mediator on Outcome}

\begin{figure}[t]
\includegraphics[width=5in]{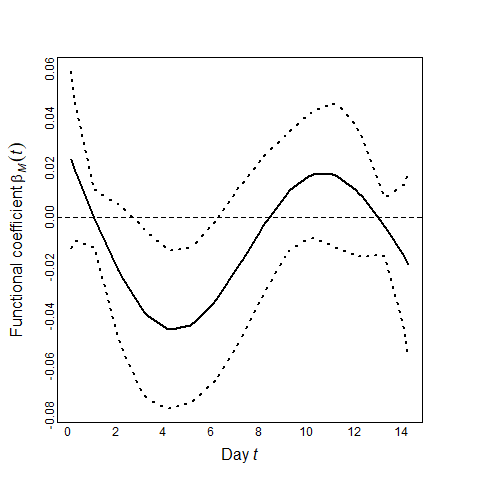} 
    \caption{ \large Functional effect of the mediator process on CO-confirmed 7-day point-prevalence abstinence 4-weeks post-TQD. Dotted lines represent confidence intervals.}
    \label{fig:funreg}
\end{figure}

The effect of the mediator on the outcome, CO-confirmed 7-day point prevalence abstinence at 4 weeks post-TQD, is shown in Figure \ref{fig:funreg}. Results from functional regression are more difficult to interpret precisely than those from a time-varying effect model \citep{Dziak2019}. Nevertheless, it is evident from the confidence intervals in Figure \ref{fig:funreg} that for the most part, the confidence intervals include 0, with the exception of the end of the first week post-TQD. All else being equal, craving in the first week (especially around day 4) has a negative weight in predicting abstinence, while craving in the second week (especially around day 10) has a weak and perhaps non-significant positive weight. A negative weight followed by a positive weight may be an indication that abstinence is more likely in individuals for whom craving increases over time. However, this explanation is intuitively unlikely because abstainers are expected to experience a decrease in craving over time, so it might be more likely that the result simply means that very high early craving can result in relapse. The wide confidence intervals also make interpretation more difficult.

\section{Discussion} \label{Sec:Discussion}

In this paper, we proposed and illustrated a method for estimating and testing the indirect effect of a randomized treatment on a distal binary outcome variable, as mediated by the nonparametric trajectory of an intensively measured longitudinal process. This continued the work of \citet{Lindquist2012}, in combining causal analysis in the counterfactual framework,  with regularized overparameterized semiparametric estimation. These two sets of ideas are among the most important in modern statistics \citep{Gelman2021}. In extending Lindquist's ideas to a binary setting, we used simulations and a real-data example to illustrate the potential use of functional mediation analysis with binary outcome data. We provide a new R package, \verb|funmediation|, for fitting this and similar models.

\subsection{Limitations}

A remaining methodological challenge in functional regression modeling approaches is the issue of missing data and possible informative dropout. If some observations per subject are missing sporadically and completely at random, this would not be a major problem. Many EMA designs give prompts at random times, or at a random number of times, so methods for analyzing them are constructed to allow non-shared measurement times and unequal coordinates. This leads to the simple approach used in this paper, which is to treat occasions of nonresponse to a prompt as though the prompt had never occurred at all on those occasions (essentially listwise deletion, although on the scale of the observation, not the entire participant). This approach assumes that values of $M(t)$ are missing completely at random, which is not ideal.

The situation is more serious if participants drop out for long periods of time, and especially if this might occur for reasons that relate to the variables of interest (e.g., because of being embarrassed or discouraged by a relapse). If the composition of the sample differs at different points in the time interval of interest, then it may be difficult to interpret the  overall trajectory. About 14\% of the participants in our analysis sample provided less than 20 observations, so even in the absence of informative missingness, their interpolated $M(t)$ trajectories may not closely match what they would have been if more data had been provided.

Multiple imputation can be beneficial for handling missing values in longitudinal data \citep[see][]{Schafer2002,Lee2016}. More research is required regarding how to use multiple imputation for functional regression \citep{ciarleglio}, which is likely to be more challenging than in the more classic case of a few waves of data at fixed times. If informative dropout is suspected, another possibility might be to jointly model the longitudinal processes together with a survival process for dropout, possibly treated as a competing risk for the main outcome of interest \citep[see][]{Liu2014}. However, this also is likely to be challenging to implement in a functional mediation context.

In addition to an overall indirect effect integrated over time, \citet{Lindquist2012} also estimated the product, $\alpha_X(t) \beta_M$(t), for a grid of times $t$, leading to a time-specific indirect effect. In principle, bootstrap tests could be done to test whether individual time-specific indirect effects are statistically significant, although Lindquist recommends a false discovery rate adjustment in this case, but this may lead to severe power limitations. We did not consider time-specific indirect effects here. Functional regression coefficients can be somewhat difficult to interpret on a time-specific basis with longitudinal data, at least without additional model assumptions. The mediator values at different time points are not independent, so that the most statistically predictive time period need not be the most causally important time period \citep[see][]{Dziak2019,Green2016,DeStavola2021,Gao2022}. Furthermore, even without intensive longitudinal data, researchers face the  challenge of time-varying confounding \citep{VanderWeele2017, DeStavola2021}. That is, if $Y$ had also been time-varying, then earlier values of $Y$ could affect later values of both $M$ and $Y$, becoming a time-varying confounder which would violate Assumption 4. We focus on testing the overall effect of the mediator process on a single later time point $Y$ in order to avoid these issues for now, although they will be important to address if functional mediation is to become more widely used for causal inference.

Similar to \citet{Lindquist2012} and \citet{ZengFanLi2021}, we assumed no treatment-mediator interaction in order to identify direct and indirect effects. \citet{Lindquist2012} does explore an alternative instrumental variables approach, which requires alternative assumptions. However, this alternative might be more complicated to extend to models that do not use the identity link. As an alternative to natural direct and indirect effects, the use of interventional \citep[see][]{VanderWeele2017} or organic \citep[see][]{Lok2020} effects might make causal inference in the presence of time-varying confounders more manageable. 

We followed \citet{Lindquist2012} in taking a frequentist approach, but a Bayesian approach to mediation analysis may also be useful \citep{Miocevic2018,KuanLiu2020}. In particular, \citet{Huang2017} described a Bayesian analogue to the time-varying mediation method of \citet{Lindquist2012} in the case where the outcome was  time-varying and normally distributed, and the ideas could presumably be adjusted to the case of a point and/or binary distal outcome.

Finally, we have not considered the possibility of periodic functions, such as diurnal or seasonal rhythms, as indexes for the varying coefficients in place of linear time. There are some settings in which such features may be theoretically and practically important \citep{Adam2017}.

\subsection{Future Directions}

Our simulation and example have considered a single time-varying variable as the mediator. Multiple time-varying mediators may be of interest, and options for such models are briefly reviewed in Appendix 2.

We have only considered scalar treatment variables. However, another option would be to have a time-varying exposure $X(t)$ and/or time-varying response $Y(t)$ in addition to the time-varying mediator $M(t)$. This was explored by \cite{zhao2018functional} who provided software to fit either concurrent or historical versions of the model. The concurrent approach could be done by fitting a TVEM model to predict the mediator, and another TVEM model to predict the response, and then multiplying the appropriate coefficients pointwise (without integration). The historical approach is able to consider lagged effects either within a sliding window \citep[see also][]{Kim2011} or for the entire past. The concurrent approach is simpler but could miss important information, by only looking at essentially cross-sectional relationships and ignoring lagged ones \citep[see][]{Maxwell2011}. An alternative, highly flexible but still somewhat parametric, approach to time-varying mediation could use differential or integral equations to represent dynamic models in continuous time \citep{Albert2018}.

For point treatment and time-varying mediator and response, \citet{Xizhen2020Arxiv} use a two-step method beginning with a TVEM-like (function-on-scalar) regression to predict $M$ from $X$, but follow with another function-on-function regression instead of a scalar-on-function regression when predicting $Y$ from $M$ and $X$. \citet{Xizhen2020Arxiv} use local polynomial estimation \citep{Fan2000} instead of splines to handle the estimation of the semiparametric time-varying effect coefficients, but either approach seems viable.

Functional mediation analysis is a novel method, relevant for both methodologists and applied researchers interested in understanding how an intensively measured longitudinal variable can mediate the relationship between treatment and study outcomes. In the presented approach, we developed functional mediation, a method for estimating a smooth, nonparametric trajectory from treatment to an intensively measured longitudinal mediator using TVEM \citep{Tan2012}, as well as estimating a smooth, nonparametric trajectory from an intensively measured longitudinal mediator to a binary end-of-study outcome using functional regression \citep{RamsaySilverman2005}. More importantly, we provided a mathematical definition of the causal direct and indirect effect that combined TVEM and functional regression in terms of potential outcomes and provided guidance on how to estimate effects using the R package \verb|funmediation| as well as how to interpret effects.

Although our approach builds on prior work examining TVEM relationships in regression \citep{Tan2012} and functional mediation \citep{Lindquist2012}, the present work is innovative in that prior work has not considered a binary, end-of-study outcome measured at only one time point. Binary end-of-study outcomes are often a primary interest in intervention studies, such as smoking cessation, weight maintenance, or abstinence from substance use at follow-up. Despite binary outcomes being a common in intervention studies, methods to evaluate the combination of intensively measured longitudinal data as well as an end of study outcome following treatment are only beginning to be developed. Functional mediation analysis, as presented here is only one of hopefully many future methods that will be developed to model nonparametric longitudinal processes as mediators to better understand how treatment impacts behavior in a causal mediation framework.



\section*{Acknowledgements}

Research was supported by National Institutes of Health grants P50 DA039838 from the National Institute of Drug Abuse and 1R01 CA229542-01 from the National Cancer Institute and the NIH Office of Behavioral and Social Science Research. Content is solely the responsibility of the authors and does not necessarily represent the official views of the funding institutions.

\section*{Declaration of interest statement}
The authors report there are no competing interests to declare.

\clearpage

\bibliography{main}

\clearpage

\section*{Appendix 1: Effect Decomposition with Log Link}

\subsection*{Proof for Log Link}

Suppose that a log link function is being used, with the effect definitions given in expression (\ref{eqnLogLinearEffects}) and the assumptions given in section \ref{Sec:Assumptions}. The mediator is continuous as given by model (\ref{eqn:M_subject}). The response is binary or count using model (\ref{eqn:Y}), with the log function for $g$. The proof follows \citet{VanderWeeleVansteelandt2010}, \citet{Lindquist2012}, and \citet{VanderWeele2015}. 

By definition, 
\begin{flalign*}
&\log(E(Y(x,\bM(\xast))|\bC=\bc))  \\
=&\log\Big(\int E\big(Y(x,M(\xast))|M(\xast)=\boldm,\bC=\bc)P(M(\xast)=\boldm|\bC=\bc) \dm \Big) 
\end{flalign*}

By the consistency and composition assumptions and assumption 4, 
we can drop the conditioning of $Y(x,\bM(\xast)$ on $\bM(\xast)$ so that 
\begin{flalign*}
&\log\Big(E(Y(x,\bM(\xast))|\bC=\bc)) \\
=&\log\Big(\int E(Y(x,\bM(\xast))|\bC=\bc,\bM(\xast)=\boldm))P(\bM(\xast)=\boldm|\bC=\bc)\dm\Big)
\end{flalign*}

By the first three assumptions, we can equate the potential outcomes with conditional expectations so that 
\begin{flalign*}
&\log(E(Y(x,\bM(\xast))) \\
=&\log\Big(\int E(Y|\bC=\bc,X=x,\bM=\boldm))P(\bM=\boldm|\bC=\bc,x=\xast)\dm \Big) 
\end{flalign*}

Then using model (\ref{eqn:Y}) in which there is no interaction between the treatment and the mediator, the above becomes  
 \begin{align}
 \label{eqnLogLinkTrick}
 &    \log\Bigg(
\int   \exp\Big( \beta_0 + \bc \bbeta_C  + \beta_X x + \int \beta(t)  \boldm(t) \dt \Big) 
P\Big(\bM=\boldm|\bC=\bc,X=\xast\Big)\dm \Bigg) 
\nonumber \\
=& 
\int \big( \beta_0 +  \bc\bbeta_C  + \beta_X x  \big)P(\bM=\boldm|\bC=\bc,X=\xast) \dm \nonumber \\
&+ 
 \int \Big( \int \beta(t) \boldm(t) \dt \Big) P(\bM=\boldm|\bC=\bc,X=\xast) \dm \\
=&  
 \beta_0 +  \bc \bbeta_C  + \beta_X x +
 E\Big( \int \beta_M(t) \boldm(t) \dt \Big | \xast \Big) \nonumber .
\end{align} 
 
Using model (\ref{eqn:M}),
 \begin{align*}
  E\big( \int \beta_M(t) \boldm(t) \dt \Big | \bc, \xast \big) &= 
\int \beta_M(t)   E\big(    \boldm(t)  | \bc, \xast \big) \dt   \\
&= \int \alpha_0 +  \bc\balpha_C(t)  +  \alpha_X(t) \xast \dt 
 \end{align*}
Therefore, 
\begin{align*}
\log(E(Y(x,\bM(\xast)|\bc)) &= 
\beta_0  + \bc \bbeta_C + \beta_X X +
 \int \beta_t
\big( \alpha_0 + \bc \balpha_C(t) + \alpha_X(t) \xast \big)
\dt.
 \end{align*}
 Now we can compute the total effect,  natural direct effect, and  natural indirect effect. First, 
\begin{align*}
    \log \TE =& \log E\Big(  Y\big(1,\bM(1)\big)\Big) - \log E\Big( Y\big(0,\bM(0)\big) \Big) \\
    &= 
\Big(
 \beta_0 + \bc \bbeta_c + \beta_X  +
 \int \beta_M(t)
\big( \alpha_0(t) + \alpha_X(t) \big)
\dt\Big)
-
\Big(
 \beta_0 + \bc \bbeta_{\changed C} + 
 \int \beta_M(t) \alpha_0(t) 
\dt\Big)
\\
&= \beta_X +
\int \alpha_X(t) \beta_M(t)
\dt.
\end{align*} 
Next,
\begin{align*}
    \log \DE =& \log E\Big(  Y\big(1,\bM(0)\big)\Big) - \log E\Big( Y\big(0,\bM(0)\big) \Big) \\
    =&  
    \Big(
 \beta_0 + \bc \bbeta_{\changed C} + \beta_X  +
 \int \beta_M(t) \alpha_X(t)
\dt
\Big) - 
\Big(
  \beta_0 + \bc \bbeta_{\changed C}  +
 \int \beta_M(t) \alpha_X(t) 
\dt
\Big) = \beta_X.
\end{align*}
Last, 
\begin{align*}
    \log \IE &= \log E\Big(  Y\big(0,\bM(1)\big)\Big) - \log E\Big( Y\big(0,\bM(0)\big) \Big) \\
    &=
    \Big(
 \beta_0 + \bc\bbeta_C + 
 \int \beta_M(t)
\big( \alpha_0 + \alpha_X(t) \big)
\dt
\Big) -
\Big(
 \beta_0 + \bc\bbeta_C + \int \beta_M(t)
 \alpha_0(t) 
\dt
\Big) \\
    &= \int \alpha_X(t) \beta_M(t) \dt.
\end{align*}

Comparing the above results, it is clear that the direct and indirect effects add (on the log scale) to the total effects, as the definitions of these effects would imply. The proof above would not work exactly for a logit link, because the equivalent of (\ref{eqnLogLinkTrick}) would not hold;  it is generally not the case that $\mathrm{logit}(a)+\mathrm{logit}(b) = \mathrm{logit}(ab)$.

A further limitation is that, following \citet{Lindquist2012}, this proof treats $\bM(t)$ as fully observed, so that it can be replaced with its expectation $\alpha_0 + \alpha_X(t) X$ without any bias.  In practice, the sample estimates do not exactly decompose, presumably because of smoothing.  
\clearpage

\section*{Appendix 2:  Extending the Model to Multiple Treatments or Multiple Mediators} \label{Sec:Extend}

So far, we did not consider the possibility of more than one mediator process.  One way to handle multiple mediators is to  study the joint mediated effect through all of the mediators
\citep[see][]{VanderWeele2015,Wang2013}, which could be estimated by subtracting the natural direct effect from the total effect. In this case, the natural direct effect represents the effect that does not go through any of the mediators. If the no-interaction assumption is made, then the controlled direct effect is assumed to equal the natural direct effect, and thus the joint mediated effect may be estimated by subtracting the controlled direct effect from the total effect. Of course, this approach does not reveal which of the mediators, individually, is most important.

Separately estimating the effects of multiple treatment variables and multiple mediator variables is also possible, although the number of parameters required increases quickly. Suppose there are $p$ treatment variables, each either dichotomous or numerical. For example, a set of $p$ binary treatment variables might represent dummy codes for a categorical treatment variable with $p+1$ levels, such as $p$ treatment groups being compared to a single control group. 

Suppose there are $q$ mediator variables. Then the model for the mediator could be specified as
\begin{equation*}
E(\bM(t)) = \bAlpha_0(t) + \bAlpha(t)\mathbf{X}
\end{equation*}
for the vector of time-varying mediators, where $\bAlpha$ is a $q_2 \times p$ matrix of functions of $t$, and the model for the outcome could be specified as
\begin{equation}
\label{eqnMultipleMediators}
g(E(Y))=\beta_0 + \sum_{k=1}^p \beta_{x_k} x_k + 
\sum_{\ell=1}^{q} \int_t \beta_{M_\ell}(t) M_\ell(t) \dt .
\end{equation} 
If mediator, $\ell$, is  not time-varying and does not have time-varying effects, the integral $\int \beta_{M_\ell}(t) M_\ell(t) \dt$ is not needed and could simply be replaced by a product, $\beta_{M_\ell} M_\ell$, for some scalar, $\beta_{M_\ell}$. Likewise, in this case, the $\ell$th column in $\mathbf{A}(t)$ would equal a constant $A_\ell$ across all observation times.

In total, if there are $p$ treatment variables and $q$ mediator variables, then there are $pq$ distinct indirect effect pathways. As a caveat, obtaining identifiable estimates and useful confidence intervals from this model might be difficult unless $p$ and $q$ are very small or the sample size is very large. This is because the underlying model for $Y$ is essentially a regression with many covariates (considerably larger than $pq$, depending on the effective number of time points after smoothing) but only one observation per subject for the outcome. One option for making this problem easier would be to impose parametric assumptions, such as linear or quadratic trajectories, for the mediator or for the $\alpha$ and $\beta$ coefficients. A decaying exponential trajectory for the effects may be more interpretable than a quadratic trajectory \citep{Fritz2014}. Alternatively, the effects of a mediator might be constrained to be exactly cumulative, which amounts to its $\beta$ being constant over time \citep[see][]{Chen2016}. 

The model shown in Equation (\ref{eqnMultipleMediators}) treats each mediator as a parallel pathway from treatment to outcome, so that mediators never interact or affect each other. If the number of mediators and number of measurement times were both very small, then it might be feasible to go beyond this restriction. One could then estimate the path-specific effects \citep{Daniel2015}. However, this will generally not be feasible or interpretable for functional data, because the problems described earlier would be increased by the large number of pathways. Models with large numbers of mediators, such as in genomics studies, might be more challenging to combine with EMA data because of the resulting multiplication of the number of parameters. However, it should be noted that functional mediation models share some similarities with high-dimensional mediation models, in that both involve some form of regularization or data reduction to make the mediator tractable \citep{Derkach2019,Zhao2020}. Therefore, future work in both areas might lead to further adaptations of ideas from one to the other.

As an alternative to using multiple mediators, it might be beneficial to use  interventional direct and indirect effects \citep{Chen2016,Vansteelandt2017}, which do not require assuming that no confounders of the mediator and outcome  have been affected by the treatment. A disadvantage is that  interventional direct and indirect effects do not necessarily sum to the total effects. The controlled direct effect is a special case of an interventional direct effect and thus, it is also identified in the presence of a mediator-outcome confounder that has been affected by the treatment. 

\clearpage

\section*{Appendix 3:  Analysis excluding participants with two or fewer EMA prompts post-TQD} \label{Sec:Sensitivity}

We reanalyzed our empirical example by excluding the 21 participants who had two or fewer EMA prompts post-TQD. The overall conclusions remain unchanged. The total effect estimate was not statistically significant (estimate = -0.016, std. error = 0.136, $z = -0.119$, $p = 0.906$). The indirect effect estimate was statistically significant (estimate = 0.043, percentile method 95\% bootstrap confidence interval = (0.005, 0.114)). The estimate of the direct effect was not statistically significant (estimate = -0.067, std. error = 0.139). The impact of excluding participants with greater than 97\% missing EMA responses is reflected in the total effect estimate but not in the direct and indirect effect estimates. This could be interpreted as the proposed technique's strength in estimating functional mediation processes in spite of missingness. However, the impact of missingness needs to be thoroughly investigated before claiming this interpretation as scientific evidence of the technique's functionality.

\clearpage

\section*{Supplementary Material: R Code}
The code below was used to fit the model to the data set \texttt{eos1}. Further details on the function arguments can be found in the vignette tutorial in the \texttt{funmediation} R package.
\begin{verbatim}
    eos1 <- read.csv("eos1.csv")
    funresult <- funmediation(data=eos1,
                           treatment=NRT1,
                           mediator=WantToSmokeLst15minC,
                           outcome=abstDay28,
                           id=SubjectID,
                           time=timeseq,
                           binary_outcome = TRUE,
                           interpolate = FALSE,
                           nboot=500)
                           
    print(funresult)
    plot(funresult, what_plot="pfr")
    plot(funresult, what_plot="coefs")
    plot(funresult, what_plot="tvem")
\end{verbatim}

\clearpage

\begin{table}[p]
\centering
\caption{ \large Simulation Results for TVEM Predicting Mediator from Treatment}
\label{Table:Results:XM}
\begin{tabular}{lrrr}
  \hline
Sample size & 250 & 500 & 1000 \\  
  \hline
  \multicolumn{4}{l}{Intercept function $\alpha_0(t)$} \\
  \hline
Bias &  -0.0010 & -0.0002 & -0.0006 \\ 
  Mean squared error & 0.0105 & 0.0054 & 0.0029 \\ 
  Root mean squared error & 0.1022 & 0.0736 & 0.0537 \\ 
  Mean estimated SE  & 0.1038 & 0.0739 & 0.0535 \\ 
  Pointwise coverage   & 0.9461 & 0.9464 & 0.9460 \\  
  Familywise coverage   & 0.6422 & 0.6270 & 0.6276 \\
  \hline
\multicolumn{4}{l}{Treatment effect function  $\alpha_X(t)$} \\
  \hline
  Bias & 0.0001 & 0.0008 & 0.0001 \\
  Mean squared error & 0.0203 & 0.0106 & 0.0057 \\
  Root mean squared error & 0.1423 & 0.1031 & 0.0758 \\ 
  Mean estimated SE & 0.1444 & 0.1032 & 0.0747 \\ 
  Pointwise coverage & 0.9485 & 0.9531 & 0.9510 \\  
  Familywise coverage & 0.7170 & 0.6956 & 0.6412 \\  
\hline
\end{tabular}
\end{table}
 
\begin{table}[p]
\centering
\caption{ \large Simulation Results for Functional Regression Predicting Outcome from Mediator}
\label{Table:Results:MY}
\begin{tabular}{lrrr}
  \hline
Sample size & 250 & 500 & 1000 \\  
  \hline
  \multicolumn{4}{l}{Intercept $\beta_0$} \\
  \hline
Bias & 0.0277 & 0.0236 & 0.0301 \\
  Mean squared error & 0.0592 & 0.0284 & 0.0145 \\
  Root mean squared error & 0.2432 & 0.1685 & 0.1203 \\ 
  Mean estimated SE & 0.2379 & 0.1668 & 0.1174 \\ 
  Coverage & 0.9492 & 0.9504 & 0.9488 \\  
    \hline  
  \multicolumn{4}{l}{Treatment $\beta_X$} \\
  \hline
Bias & -0.0161 & -0.0176 & -0.0241 \\ 
  Mean squared error  & 0.0835 & 0.0410 & 0.0203 \\ 
  Root mean squared error & 0.2889 & 0.2024 & 0.1424 \\ 
  Mean estimated SE & 0.2833 & 0.1986 & 0.1399 \\
  Coverage  & 0.9486 & 0.9434 & 0.9504 \\ 
    \hline  
\multicolumn{4}{l}{Mediator conditional effect function $\beta_M(t)$ } \\
  \hline
  Bias  & -0.0186 & -0.0180 & -0.0219 \\ 
  Mean squared error & 0.1329 & 0.0651 & 0.0339 \\
  Root mean squared error & 0.3645 & 0.2552 & 0.1841 \\ 
  Pointwise coverage & 0.9603 & 0.9577 & 0.9552 \\ 
  Familywise coverage & 0.8598 & 0.8492 & 0.8364 \\ 
   \hline
\end{tabular}
\end{table} 

\begin{table}[tpb]
\centering
\caption{ \large Simulation Results for Indirect Effect}
\label{Table:Results:XY}
\begin{tabular}{lrrr}
  \hline
Sample size & 250 & 500 & 1000 \\ 
  \hline
  \multicolumn{4}{l}{Performance of point estimate} \\
    \hline
Bias & 0.0220 & 0.0165 & 0.0160 \\  
Mean squared error & 0.0140 & 0.0070 & 0.0036 \\ 
Root mean squared error & 0.1182 & 0.0839 & 0.0603 \\
Mean estimated SE & 0.1325 & 0.0869 & 0.0593 \\  
\hline
  \multicolumn{4}{l}{Bootstrap coverage by method} \\
    \hline
  Normal & 0.9648 & 0.9500 & 0.9340 \\ 
  Basic & 0.9682 & 0.9516 & 0.9348 \\ 
  Percentile & 0.9576 & 0.9512 & 0.9452 \\ 
  \hline
  \multicolumn{4}{l}{Bootstrap power by method} \\
    \hline
  Normal & 0.2366 & 0.6350 & 0.9284 \\ 
  Basic  & 0.1570 & 0.5654 & 0.9104 \\
  Percentile & 0.3704 & 0.6906 & 0.9420 \\ 
   \hline 
\end{tabular}
\end{table} 

\clearpage

\clearpage

\clearpage

\clearpage

\clearpage

\section*{Figure Captions}

\begin{enumerate}
\item Simple (left) versus functional (right) mediation with single-time treatment and outcome.
\item Mediation model with functional $M$, including baseline covariates $\boldsymbol{C}$.
\item Time-varying effect of varenicline (vs. nicotine patch only) on the mediator, craving to smoke in the last 15 min. Left and right panes represent time-varying intercept function and time-varying effect of treatment, respectively. Upper and lower lines represent pointwise approximate 95\% confidence intervals.
\item Functional effect of the mediator process on CO-confirmed 7-day point-prevalence abstinence 4-weeks post-TQD. Dotted lines represent confidence intervals.
\end{enumerate}
 
\end{document}